\documentclass[preprint,3p,times,12pt]{elsarticle}

\usepackage{lmodern}
\usepackage{hyperref}
\usepackage{amssymb}
\usepackage{amsmath}
\usepackage{setspace}
\usepackage{xcolor}
\usepackage{caption}
\usepackage{booktabs}
\usepackage{array}
\usepackage{longtable}
\newcolumntype{P}[1]{>{\raggedright\arraybackslash}p{#1}} % ragged-right p-column

\usepackage[numbers]{natbib}

\journal{Elsevier}

\begin{document}
%============================================================

\begin{frontmatter}

\title{Towards 6G Intelligence: The Role of Generative AI in Future Wireless Networks}

\author[stanford]{Muhammad Ahmed Mohsin}
\ead{muahmed@stanford.edu}
\author[nust]{Junaid Ahmad}
\ead{jahmad.bee20seecs@seecs.edu.pk}
\author[nust]{Muhammad Hamza Nawaz}
\ead{mnawaz.bee20seecs@seecs.edu.pk}
\author[glasgow]{Muhammad Ali Jamshed}
\ead{muhammadali.jamshed@glasgow.ac.uk}

\affiliation[stanford]{School of Electrical Engineering, Stanford University, Stanford, CA 94305, USA}
\affiliation[nust]{organization={School of Electrical Engineering and Computer Science (SEECS),\\
National University of Sciences and Technology (NUST)}, country={Pakistan}}
\affiliation[glasgow]{organization={College of Science and Engineering, University of Glasgow}, country={UK}}

% -----------------------------------------------------------
\begin{abstract}
Ambient intelligence (AmI) is a computing paradigm in which physical environments are embedded with sensing, computation, and communication so they can perceive people and context, decide appropriate actions, and respond autonomously. Realizing AmI at global scale requires sixth generation (6G) wireless networks with capabilities for real time perception, reasoning, and action aligned with human behavior and mobility patterns. We argue that \emph{Generative Artificial Intelligence} (GenAI) is the creative core of such environments. Unlike traditional AI, GenAI learns data distributions and can generate realistic samples, making it well suited to close key AmI gaps, including generating synthetic sensor and channel data in under observed areas, translating user intent into compact, semantic messages, predicting future network conditions for proactive control, and updating digital twins without compromising privacy.

This chapter reviews foundational GenAI models, GANs, VAEs, diffusion models, and generative transformers, and connects them to practical AmI use cases, including spectrum sharing, ultra reliable low latency communication, intelligent security, and context aware digital twins. We also examine how 6G enablers, such as edge and fog computing, IoT device swarms, intelligent reflecting surfaces (IRS), and non terrestrial networks, can host or accelerate distributed GenAI. Finally, we outline open challenges in energy efficient on device training, trustworthy synthetic data, federated generative learning, and AmI specific standardization. We show that GenAI is not a peripheral addition, but a foundational element for transforming 6G from a faster network into an ambient intelligent ecosystem.
\end{abstract}

% -----------------------------------------------------------
\begin{keyword}
Ambient Intelligence \sep Generative AI \sep 6G \sep Semantic
Communications \sep Digital Twins \sep GANs \sep VAEs \sep Diffusion Models
\end{keyword}

\end{frontmatter}

%============================================================
\section{Introduction}

The vision of \emph{Ambient Intelligence} (AmI) has evolved over the past two decades to depict a future in which digital environments are capable of sensing, interpreting, and proactively responding to human presence and needs. These environments, ranging from smart homes and workplaces to entire urban ecosystems, aim to seamlessly embed computational intelligence within the physical world, making the environment itself an unobtrusive yet intelligent assistant \cite{inbook}. Realizing this vision at scale demands not only a dense network of embedded sensors and actuators, but also intelligent connectivity that is responsive, adaptive, and capable of cognition. \emph{Table~\ref{tab:enablers}} provides a compact overview of the core enablers we build on in this chapter, mapping each technology to its role in AmI, the 6G specific advantage it unlocks, and a concrete example that we elaborate in later sections.

To meet these demands, sixth generation (6G) wireless networks are envisioned as more than enhanced data pipelines, they are projected as intelligent ecosystems with native support for real time perception, semantic reasoning, and distributed actuation aligned with human spatial temporal dynamics \cite{Saad2020Vision6G}. With anticipated features such as extreme data rates, ultra low latency, high reliability, massive device connectivity, and integrated sensing and communication capabilities, 6G stands as the ideal substrate for enabling ambient intelligent services \cite{Dang2020What6G}.

However, traditional AI methods alone may not suffice to realize such autonomous and context aware systems. \emph{Generative Artificial Intelligence} (GenAI), a class of models capable of learning data distributions and synthesizing realistic content, introduces a transformative shift in how intelligence is embedded into wireless environments. GenAI’s ability to generate novel, high fidelity data makes it especially suitable for addressing core AmI challenges, including synthesizing missing sensor or wireless channel data \cite{8663987}, compressing user intent into semantically rich low bit messages \cite{celik2024dawn}, predicting network states for proactive resource allocation \cite{zhong2024enabling}, and powering privacy preserving digital twin updates \cite{tao2024wireless,Bashir2023MINDTwin}.

In this chapter, titled \emph{Towards 6G Intelligence: The Role of Generative AI in Future Wireless Networks}, we argue that GenAI is not merely a supporting tool, but a foundational pillar for scalable, distributed, and adaptive AmI systems. We first examine foundational GenAI architectures, including Generative Adversarial Networks (GANs), Variational Autoencoders (VAEs), diffusion models, and generative transformers, and evaluate their suitability across different AmI application domains, such as spectrum co existence, ultra reliable low latency communication (URLLC), resilient security frameworks, and rich digital twin modeling \cite{xu2025decentralization}.

We also survey enabling technologies within the 6G landscape that can accelerate and host these generative models. These include distributed computing paradigms like edge and fog computing, massive IoT device swarms, intelligent reflecting surfaces (IRS), and non terrestrial networks (NTNs) \cite{cui2025overview}. Such infrastructures are crucial for supporting decentralized GenAI frameworks and for reducing inference latency at the edge.

Furthermore, we discuss emerging challenges that must be addressed to fully integrate GenAI into ambient intelligent 6G systems. These include the development of energy efficient on device training strategies, ensuring the trustworthiness and generalizability of synthetic data, enabling federated and privacy preserving generative learning, and creating AmI specific standardization protocols \cite{khoramnejad2024generative}.

Ultimately, we demonstrate that GenAI represents a paradigm shift, from reactive AI pipelines to proactive, generative ecosystems, positioning 6G not just as a faster network but as the intelligence fabric underpinning the ambient aware world of tomorrow.

\renewcommand{\arraystretch}{1.3} % increases row height
\begin{table}[!t]  % top-of-page placement
\centering
\small
\caption{Key Enablers of Ambient Intelligence in 6G}
\label{tab:enablers}
\begin{tabular}{p{3cm} p{3cm} p{6cm} p{3cm}}
\hline
\textbf{Enabler} & \textbf{Role in AmI} & \textbf{6G-Specific Advantage} & \textbf{Example} \\
\hline
Generative AI & Context-aware decision-making & Low-latency, edge-deployed models & Real-time health monitoring \\

Massive IoT & Rich environmental sensing & High-density device connectivity & Smart factories \\

THz Communication & High data throughput & Supports ultra-high-resolution sensing & Immersive AR/VR \\

Digital Twins & Predictive simulation & Near real-time synchronization & Energy-efficient smart grids \\

Edge Intelligence & Localized processing & Reduced backhaul load & Smart traffic systems \\
\end{tabular}
\end{table}
\renewcommand{\arraystretch}{1} % reset back

%============================================================
\section{Background}

% --- Table first (top of section) ---

Ambient Intelligence (AmI) envisions environments enriched with pervasive sensing, adaptive decision making, and seamless human–machine interaction. Its realization depends on the convergence of artificial intelligence and next generation wireless networks. Within this vision, 6G is not merely a communication enabler, it is the connective tissue for context aware, self optimizing ecosystems. Foundational studies such as Helin et al. outline AI enabled architectural frameworks for adaptive network management, automatic service provisioning, and intelligent orchestration across diverse layers of communication infrastructure \cite{yang2019aienabled}. Table 2 distills the recurring radio–compute pressures in AmI deployments and aligns them with four 6G enablers, edge and fog computing, IoT swarms, intelligent reflecting surfaces, and non terrestrial networks, while the final column highlights the specific GenAI lever that converts scarce measurements into priors, produces short horizon forecasts, and enables privacy preserving updates.

A key enabler of AmI is \emph{edge AI}, where computation and learning are shifted closer to devices and sensors, minimizing latency, enhancing privacy, and enabling rapid adaptation \cite{letaief2021edge}. Beyond edge intelligence, the emergence of wireless Big AI Models (wBAIMs) offers the promise of few shot adaptability across a wide range of wireless and contextual tasks. Such models align closely with the needs of AmI, where heterogeneous and dynamic contexts demand generalizable and continuously learning intelligence \cite{chen2023big}.

Generative AI (GenAI) further expands these possibilities, acting not only as a predictive tool but as a creative mechanism for simulating scenarios, synthesizing multimodal sensor data, and modeling latent human–environment interactions. Khoramnejad and Hossain highlight GenAI’s use in optimizing wireless networks through proactive planning, resource allocation, and diffusion based modeling in dynamic and non terrestrial environments, all of which contribute to the responsiveness and adaptability expected in ambient intelligent systems \cite{khoram2024generative}. 

Recent work also emphasizes the role of generative models in constructing \emph{digital twins} of wireless environments. Tao et al. propose hierarchical generative AI enabled twins that replicate communication behaviors at both the message and policy level, enabling real time synchronization, network slicing, and emulation of complex user–environment interactions \cite{tao2024wireless}. Similarly, security oriented applications of GenAI illustrate how AmI environments may self diagnose vulnerabilities, simulate attack strategies, and design adaptive defense frameworks \cite{yang2025vulnerability}. 

\begin{figure}[!t]
    \centering
    \includegraphics[width=0.6\textwidth]{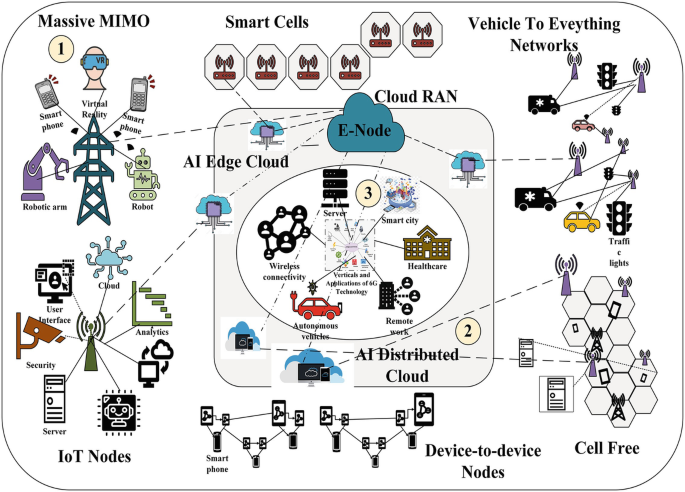}
    \caption{An ecosystem showing how sensors, communication infrastructure, computation nodes, and AI interact in real time to enable Ambient Intelligence. Generative AI augments this loop by synthesizing adaptive responses, predicting user or system needs, and generating optimized strategies.}
    \label{fig:ambient_ai_ecosystem}
\end{figure}

Cui et al. provide a broader perspective on AI–6G integration by categorizing developmental trajectories into “AI for network,” “network for AI,” and “AI as a service,” highlighting standardization, efficiency, and scalability as critical challenges \cite{cui2025overview}. Within AmI contexts, these trajectories translate into practical requirements for robustness, seamless integration, and human centric trust.

Together, this body of work underscores a rapid transition, from reactive intelligence limited to task specific optimizations, toward proactive, generative, and adaptive ecosystems that embody the essence of ambient intelligence.
%============================================================

{
\scriptsize
\renewcommand{\arraystretch}{1.01}
\setlength{\tabcolsep}{1pt}

\begin{longtable}{P{0.15\textwidth} P{0.15\textwidth} P{0.15\textwidth} P{0.15\textwidth} P{0.15\textwidth} P{0.15\textwidth}}
\caption{Design challenges and GenAI enabled levers across edge and fog computing, IoT device swarms, intelligent reflecting surfaces, non terrestrial networks, and their integrated operation.}
\label{tab:enablers-challenges}\\
\toprule
\textbf{Challenge area} &
\textbf{Edge/fog} &
\textbf{IoT swarms} &
\textbf{IRS} &
\textbf{NTN} &
\textbf{GenAI integration} \\
\midrule
\endfirsthead

\toprule
\textbf{Challenge area} &
\textbf{Edge/fog} &
\textbf{IoT swarms} &
\textbf{IRS} &
\textbf{NTN} &
\textbf{GenAI integration} \\
\midrule
\endhead

\midrule
\multicolumn{6}{r}{\emph{Continued on next page}}\\
\midrule
\endfoot

\bottomrule
\endlastfoot

\textbf{Coverage} &
Micro data centers near sensors shorten paths for ambient services. &
Dense devices widen sensing, rural and indoor gaps remain. &
Programmable reflections raise line of sight likelihood with careful siting. &
Space and stratosphere extend reach beyond terrestrial cells with tight link budgets. &
Digital twins with generative channels synthesize radio maps for placement and surfaces. \\

\textbf{Latency/jitter} &
Close to data execution lowers loop delay for perception, reasoning, and actuation. &
Local loops are fast, sleep and duty cycles interrupt timing. &
Shaped multipath reduces delay spread, controller updates add reconfiguration delay. &
Propagation and beam reselection add tens of milliseconds and jitter. &
Short horizon generative predictors prefetch features and choose low jitter routes. \\

\textbf{Energy/power} &
Edge training and adaptation face thermal and power caps at sites. &
Battery nodes rely on event driven sensing and low power modes. &
Mostly passive surfaces, controller and control link overhead still matter. &
Air and space payloads are power limited for compute and links. &
Energy aware finetuning, compact samplers, and policy search balance budgets. \\

\textbf{Spectrum} &
Busy bands near edges create co channel interference among tenants. &
Many short packets contend and collide under bursty sensing. &
Reflections steer energy to reshape spatial reuse and reduce leakage. &
Beams and gateways coordinate frequency and polarization planning. &
Generators learn interference structure and synthesize rare collisions for control. \\

\textbf{Mobility and handover} &
Stateful services need session continuity across edge domains. &
Devices join and leave frequently, participation is intermittent. &
Phase maps must anticipate user motion and moving blockers. &
Low Earth orbit beams move quickly, predictive handover is required. &
Sequence models forecast mobility to pre warm models and pre configure surfaces and beams. \\

\textbf{Backhaul} &
Model exchange and telemetry can bottleneck constrained backhaul. &
Uplink bursts from many clients congest access and gateways. &
Surfaces avoid heavy backhaul, control signaling must scale. &
Feeder links share capacity and face weather variability. &
Split learning and adaptive offload place encoders near data and keep heavy decoders in the cloud. \\

\textbf{Privacy and governance} &
Keeping raw data local aids privacy, auditing is required. &
Sensor data raises consent and retention issues in deployments. &
Surfaces change exposure patterns without processing content. &
Cross border storage and downlinks require policy alignment. &
Differential privacy and governance metadata travel with models and synthetic datasets. \\

\textbf{Federated learning} &
Edges aggregate models under strict uplink budgets and delays. &
Clients are intermittent and non identical in distribution. &
Surface context can be included for personalization and control. &
Long delays and visibility windows complicate rounds. &
Compressed updates, over the air aggregation, and split learning stabilize generative training. \\
\end{longtable}
}
\section{Generative AI Architectures}

Generative AI (GenAI) models form the algorithmic backbone of intelligent content synthesis, enabling machines not only to analyze the world but to create realistic, contextually relevant data. Unlike discriminative models, which specialize in classification or regression—generative models learn the joint probability distribution of data, allowing them to produce entirely new samples that preserve structural, semantic, and statistical properties of the original domain.

In the context of 6G-powered Ambient Intelligence, this ability is pivotal. AmI systems rely on continuous perception, anticipation, and adaptation to user needs and environmental changes. However, real-world wireless, sensing, and human–environment interaction datasets are often incomplete, noisy, or costly to collect. Generative models bridge this gap by synthesizing missing data, predicting future states, and producing realistic multimodal content that enhances situational awareness, supports decision-making, and enables autonomous operation.

Below, we examine four major families of generative architectures: GANs, VAEs, diffusion models, and generative transformers, highlighting their working principles, unique advantages, and specific roles in enabling scalable, reliable, and adaptive AmI within the 6G ecosystem.

\subsection{Generative Adversarial Networks (GANs)}

Introduced by Goodfellow et al. in 2014, Generative Adversarial Networks (GANs) pit two neural networks, a generator $G$ and a discriminator $D$,against each other in a minimax game. The generator attempts to map random noise $z \sim p_z$ into realistic samples $G(z)$, while the discriminator learns to distinguish $G(z)$ from true data $x \sim p_{data}$ by outputting a probability $D(x)$ of ``realness.'' Over successive iterations, $G$ becomes capable of producing high–fidelity samples that $D$ can no longer reliably distinguish from the real distribution \cite{goodfellow2014generativeadversarialnetworks}. 

Within the broader vision of Ambient Intelligence (AmI), GANs provide the foundation for generating highly realistic and context-aware synthetic data that enables adaptive and intelligent system behavior. For example, Qiao et al. demonstrate text-to-image generation by conditioning the generator on semantic priors, achieving photorealistic outputs from abstract prompts \cite{NEURIPS2019_d18f655c} and GANs have also been used for zero-shot learning in scenarios where data is scarce \cite{Mohsin2023FIT}. In AmI environments, such text-to-image pipelines can enhance human-computer interaction by allowing natural language prompts to generate personalized visual content, supporting domains like smart healthcare monitoring or immersive learning.

In wireless research, Zhang et al. employ a specialized ChannelGAN to learn 6G IIoT channel impulse responses directly from measurements, enabling synthetic augmentation of scarce propagation data for robust link design \cite{10965760}. This ability is vital in AmI-enabled smart factories and industrial setups, where diverse and adaptive channel conditions must be modeled without exhaustive measurements, ensuring ultra-reliable and low-latency communication (URLLC).

\begin{figure}[htbp]
  \centering
  \includegraphics[width=0.68\textwidth]{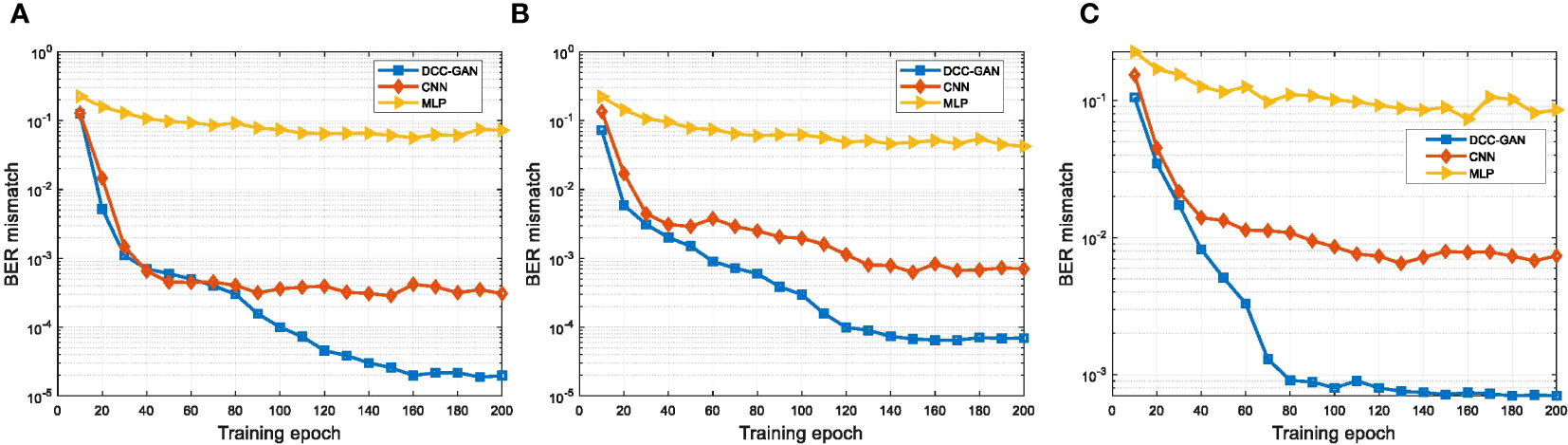}
  \caption{Bit-Error-Rate (BER) mismatch (\%) across channel emulators. The GAN-based approach yields minimal BER discrepancy, enabling reliability-sensitive Ambient Intelligence services (e.g., healthcare, industrial robotics) under dynamic conditions \cite{10.3389/fmars.2023.1149895}}
  \label{fig:ber_mismatch_gan}
\end{figure}

From a resilience perspective, GANs also support AmI security. Di Mattia et al. survey how GANs can simulate rare or adversarial anomalies, ranging from hardware faults to protocol exploits, thereby providing rich testbeds for intrusion detection systems \cite{dimattia2021surveygansanomalydetection}. In AmI-driven critical infrastructures, where systems must continuously adapt to threats without service disruption, such generative adversarial testbeds help design robust anomaly detectors.

Meanwhile, Ring et al. apply GANs to network traffic generation, using flow-based features to capture temporal dependencies across sessions. Their GAN-Net produces synthetic traces nearly indistinguishable from real enterprise traffic, which is invaluable for training and stress-testing intrusion detection and QoS controllers without compromising user privacy  \cite{RING2019156}. In an AmI context, this allows the creation of realistic digital twins of communication networks that can be continuously tested and optimized without interfering with real users.

Together, these works illustrate how the adversarial paradigm of GANs provides an adaptable framework to model complex, multimodal distributions, whether in the visual, spectral, or behavioral domains as is apparent from the figure 2. By enabling context-aware synthesis, anomaly modeling, and secure testing environments, GANs emerge as a cornerstone of generative intelligence within 6G-enabled Ambient Intelligence ecosystems.

\subsection{Variational Autoencoders (VAEs)}

Variational Autoencoders (VAEs) learn a probabilistic latent representation that can both explain observed data and generate realistic new samples. An encoder \(q_{\phi}(z \mid x)\) approximates the posterior over latent variables, while a decoder \(p_{\theta}(x \mid z)\) reconstructs the input from the latent code. Training maximizes the evidence lower bound (ELBO) \cite{kingma2013auto,rezende2014stochastic}:
\[
\mathcal{L}(\theta,\phi;x)
= \mathbb{E}_{z \sim q_{\phi}(z\mid x)} \bigl[ \log p_{\theta}(x \mid z) \bigr]
\;-\;
\mathrm{KL}\!\bigl(q_{\phi}(z\mid x)\,\|\,p(z)\bigr),
\]
balancing a reconstruction term with a Kullback–Leibler regularizer that shapes the latent space, typically toward a standard Gaussian prior. The reparameterization trick enables low-variance gradient estimates, allowing end-to-end training with stochastic latent variables \cite{kingma2013auto,rezende2014stochastic}. By encoding sensory input into a structured latent space and decoding it back into the data domain, VAEs enable robust representation learning while capturing uncertainty. This makes them particularly well-suited for Ambient Intelligence (AmI), where systems must process multimodal, noisy, and context-dependent data streams in real time.

In Ambient Intelligence (AmI) environments, Variational Autoencoders (VAEs) serve as powerful priors and semantic compressors for both perception and communication tasks. Baur \emph{et al.} demonstrate this by training a VAE directly on noisy channel observations and then using its learned generative prior to parameterize a near-MMSE channel estimator; evaluated on real measurement data, the VAE-based estimator surpasses strong learning baselines and benefits from synthetic-data pretraining, illustrating that VAEs can capture complex radio-environment structures even without perfect CSI labels \cite{baur2023vaeCE}. Such capabilities are crucial in AmI settings, where context-aware devices must operate under incomplete or noisy observations while still maintaining reliable connectivity.  

For end-to-end digital semantic communication, Bo \emph{et al.} design a VAE-based joint coding–modulation scheme that learns discrete constellations consistent with channel conditions while preserving digital implementation practicality. Their model maintains semantic fidelity across SNRs and rates by optimizing a VAE objective tailored to source–channel compatibility \cite{bo2023jcmvae}. In AmI systems, this type of semantic preservation ensures that user intent, sensor data, or contextual cues can be transmitted without degradation, even in bandwidth-limited or interference-prone environments.  

\begin{figure}[t]
  \centering
  \includegraphics[width=0.72\textwidth]{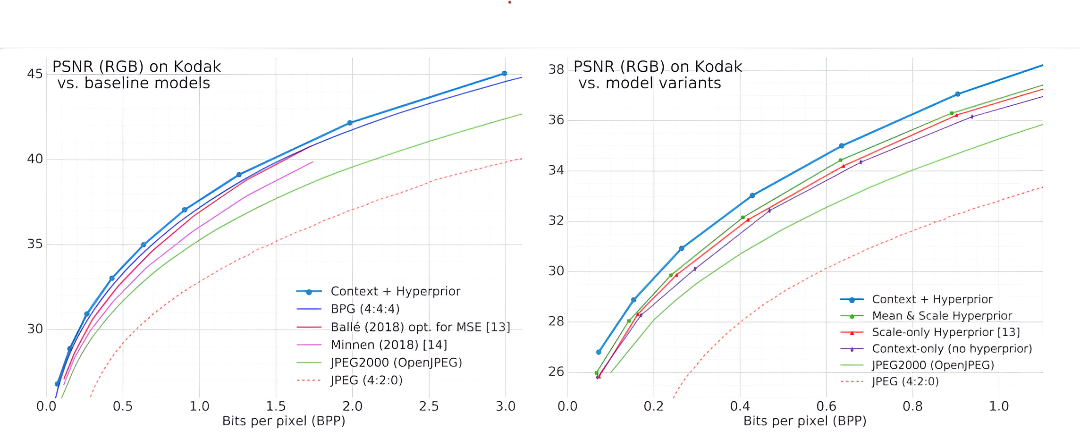}
  \caption{Rate–distortion trade-off for learned variational compression (context + hyperprior) compared with standard codecs on the Kodak dataset. The VAE-based model attains lower distortion at the same bitrate compared with JPEG, JPEG2000, and BPG. Reproduced from \cite{minnen2018joint} (Fig.~2).}
  \label{fig:vae_rate_distortion_ami}
\end{figure}

To reduce downlink overhead in massive MIMO FDD systems, Shin \emph{et al.} propose a vector-quantized VAE with shape–gain quantization and trainable Grassmannian codebooks, yielding improved CSI reconstruction at fixed feedback budgets and lowering quantization complexity \cite{shin2024vqvae}. These mechanisms translate naturally into AmI infrastructures, where distributed IoT devices and intelligent surfaces must efficiently exchange environmental state information while keeping feedback lightweight and adaptive.  

Under high-mobility, doubly selective channels, Li \emph{et al.} show that a VAE-based receiver for OTFS can jointly handle channel estimation and symbol detection in the delay–Doppler domain without dedicated pilots by optimizing an ELBO-derived objective \cite{li2021otfsvae}. This robustness to severe time–frequency dispersion is particularly valuable in AmI applications such as smart transportation or human activity monitoring, where mobility and multipath are inherent.  

Taken together, these studies demonstrate how VAEs’ structured latent spaces and principled uncertainty handling enable semantic compression, robust estimation from noisy or partial observations, and data-efficient adaptation, capabilities that are foundational for scalable and resilient Ambient Intelligence ecosystems.

\subsection{Diffusion Models}

Diffusion probabilistic models generate data by reversing a gradual noising process. A \emph{forward} Markov chain progressively corrupts data with Gaussian noise,
\[
q(\mathbf{x}_t \mid \mathbf{x}_{t-1}) = \mathcal{N}\!\left(\sqrt{1-\beta_t}\,\mathbf{x}_{t-1},\, \beta_t \mathbf{I}\right), \quad t=1,\dots,T,
\]
with a variance schedule $\{\beta_t\}$. The \emph{reverse} chain is parameterized to denoise step by step back to $\mathbf{x}_0$. Ho \emph{et al.} showed that training can be formulated as predicting the injected noise directly, leading to a simple mean–squared error objective
\[
\mathcal{L}_\text{simple} = \mathbb{E}_{t,\mathbf{x}_0,\boldsymbol{\varepsilon}}
\bigl[\lVert \boldsymbol{\varepsilon} - \boldsymbol{\varepsilon}_\theta(\mathbf{x}_t,t)\rVert_2^2\bigr],
\]
which provides stable optimization and high–fidelity synthesis \cite{ho2020ddpm}. Earlier, Sohl-Dickstein \emph{et al.} introduced the forward–reverse non-equilibrium thermodynamics interpretation that underpins modern diffusion training \cite{sohldickstein2015noneq}. Subsequent refinements in noise schedules and loss weighting further improved sample quality \cite{nichol2021improved}, while diffusion was shown to outperform GANs in both diversity and fidelity of samples \cite{dhariwal2021beatgans}.

An alternative formulation is \emph{score-based generative modeling}, where one learns the score $\nabla_{\mathbf{x}}\log p_t(\mathbf{x})$ and simulates a reverse-time stochastic differential equation (SDE) to generate samples \cite{song2019score,song2021sde}. This approach offers principled control of noise dynamics, predictor–corrector samplers, and effective conditional generation.

\begin{figure}[htbp]
  \centering
  \includegraphics[width=0.8\textwidth]{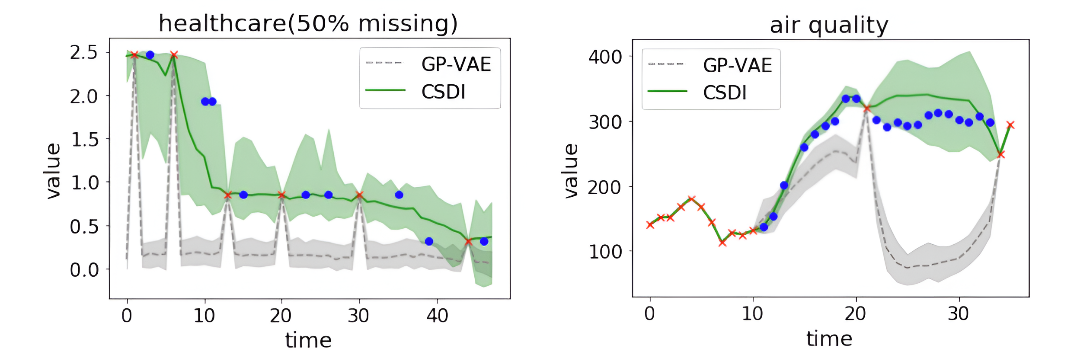}
  \caption{ Examples of probabilistic time series imputation on healthcare (left) and air quality (right) datasets. Red crosses denote observed values, blue circles mark ground-truth imputation targets, and shaded regions show the 5\%–95\% quantiles of the model’s predictions. Reproduced from Tashiro \emph{et~al.} \cite{tashiro2021csdi}.}
  \label{fig:genai_imputation}
\end{figure}

Diffusion-based architectures, can enhance the resilience of Ambient Intelligence by reconstructing missing or corrupted sensor streams. Figure~\ref{fig:genai_imputation} illustrates a sensor time-series where a diffusion-like model performs imputation compared to a baseline approach. While the baseline achieves slightly lower mean absolute error, the generative method offers smoother reconstructions and captures the underlying dynamics more robustly. Such capabilities are particularly valuable for health monitoring, industrial IoT, and smart city deployments where sensor reliability cannot always be guaranteed.

For Ambient Intelligence, these properties are highly desirable. Diffusion models can synthesize realistic sensor streams (e.g., IoT data, ambient sound, mobility traces) to augment training for recognition systems, model wireless traffic patterns under varying loads, or impute missing context in multimodal environments. Unlike adversarial methods, they ensure stable training without mode collapse, which is critical for safety-critical AmI applications. Conditional diffusion frameworks also allow controlled generation based on side information (e.g., partial channel state, user activity, or environmental metadata), aligning well with adaptive and context-aware 6G systems. 

Recent innovations such as classifier-free guidance \cite{ho2022cfg} and efficient solver architectures \cite{karras2022elucidating} further enhance controllability and scalability. Thus, diffusion modeling provides a compelling foundation for AmI pipelines, supporting reliable, diverse, and contextually-aware data synthesis across sensing, communication, and decision-making layers.

\subsection{Generative Transformers and Large Language Models (LLMs)}

Transformers replace recurrence with self-attention, enabling parallel sequence modeling and long-range dependency capture with superior scaling properties \cite{vaswani2017attention}. Generative pretraining turns this architecture into an autoregressive next-token predictor that, when scaled, exhibits strong zero- and few-shot generalization, in-context learning, and broad transfer with minimal or no task-specific finetuning \cite{brown2020gpt3}. Post-training alignment—via instruction tuning and reinforcement learning from human feedback (RLHF)—makes such models more helpful and reliable for open-ended tasks \cite{ouyang2022instructgpt}, while larger multimodal variants extend beyond text to accept images, audio, video, and sensor inputs \cite{openai2023gpt4}. Open foundation models (e.g., LLaMA~2) have further accelerated domain adaptation through efficient finetuning and retrieval-augmented generation \cite{touvron2023llama2}.  

Within Ambient Intelligence (AmI), these generative transformers and LLMs serve as powerful priors and adaptive planners that can reason over heterogeneous contextual signals—ranging from environmental sensors and wearable data to social interaction cues and spatial information. They enable knowledge-grounded reasoning, agentic control loops, and cross-modal orchestration of services \cite{boateng2024llmsurvey,liang2025llmwireless}. Proof-of-concept studies already demonstrate how LLMs can suggest adaptive policies in smart homes, guide personalized recommendations, and integrate with lightweight heuristics to improve safety, efficiency, and comfort \cite{lee2024llmra}.  Recent domain-specific frameworks show how RAG with multi-modal LLMs can ground wireless knowledge and telemetry for control \cite{Mohsin2025ICC}.

At the perception layer, transformer variants tailored to high-dimensional sensory streams are particularly effective. For example, Swin-Transformer–based encoders can capture long-range structure in visual or environmental data and compress contextual feedback for efficient adaptation \cite{cheng2024swincsi}. Similarly, multimodal transformers fuse inputs such as video, LiDAR, radar, and biosignals to construct a holistic context representation, enabling more accurate predictions of user intent and environment-aware decision-making \cite{tian2023multimodalxfmr}. Broader transformer advances in structure recognition, e.g., table understanding with hybrid convolution–transformer backbones, underscore the architecture’s adaptability across sensing modalities foundational to AmI \cite{10.1007/978-3-031-41734-4_26}.Beyond static states, GPT-style temporal transformers provide strong priors for forecasting user activity patterns, resource demand, and environmental dynamics, complementing diffusion- or VAE-based generative models that enrich AmI simulations and decision pipelines \cite{cao2023tempo}.  

In short, generative transformers and LLMs act as \emph{foundation planners} for Ambient Intelligence: they compress human intent, incorporate side information through prompting or retrieval, and synthesize adaptive policies that coordinate sensing, communication, and computation. When deployed with edge execution and grounded tool use, they enable ambient-intelligent behaviors that continuously adapt across contexts with minimal supervision.  
% ===============================
\section{6G Enablers for Distributed GenAI}
% ===============================

Ambient intelligent services depend on compute that is close to data, low delay communication, and fast adaptation of models to local context. The 6G stack provides these capabilities through a set of complementary enablers that can host or accelerate distributed GenAI, namely edge and fog computing, IoT device swarms, intelligent reflecting surfaces, and non terrestrial networks. In combination, they reduce end to end latency, enhance privacy by keeping sensitive observations local, and enable scalable deployment of generative models across space and time \cite{letaief2021edge}. This section introduces each enabler and explains how it supports training, inference, and orchestration of GenAI across the network continuum.

\subsection{Edge and Fog Computing for Distributed GenAI}

Edge and fog computing move computation, storage, and analytics from centralized clouds toward base stations, access points, road side units, and micro data centers, creating a continuum from device to edge to cloud \cite{bonomi2012fog,mao2017mecsurvey,etsi2015mec}. Placing generative models near data sources shortens control loops for ambient intelligence and reduces backhaul load, while also improving privacy by keeping raw sensor traces local \cite{letaief2021edge}.

At the systems level, three design patterns are especially relevant for GenAI at the edge. First, \emph{collaborative partitioning} splits a model or pipeline between device, edge, and cloud to meet latency and energy targets. Classic results such as Neurosurgeon show that layer level partitioning of deep networks lowers end to end latency and device energy substantially by selecting split points based on per layer compute and activation sizes \cite{kang2017neurosurgeon}. The same principle applies to diffusion and transformer generators, where encoders, latent transforms, or early denoisers can run at the edge, and heavier stages complete in the cloud when needed.

Second, \emph{federated and split training} enable on site adaptation of generative models without centralizing data. Federated learning coordinates many clients to learn shared parameters from locally observed distributions, which is well matched to AmI settings with heterogeneous, non identical users and environments \cite{kairouz2021fl}. For GenAI specifically, edge sites can fine tune low rank adapters for large generators or discriminators, avoiding full model updates and reducing communication \cite{hu2021lora}. Split learning variants further reduce device memory by backpropagating only through a shallow prefix locally and the remainder at the edge.

Third, \emph{fast sampling and efficient fine tuning} make modern generators practical under tight edge budgets. For diffusion models, progressive distillation and consistency model families compress many step samplers into few step or single step generators, trading a small quality loss for large latency gains \cite{salimans2022progressive,song2023consistency,heek2024mscm}. In parallel, low rank adaptation provides small parameter update heads for generative transformers and diffusion backbones, enabling site specific personalization with limited memory and bandwidth \cite{hu2021lora}.

Recent work on \emph{mobile edge generation} demonstrates cooperative text to image generation between an edge server and a user device using a latent diffusion backbone, dynamically allocating denoising steps across the link to meet delay targets and device limits \cite{zhong2024meg}. Together, these techniques show how edge and fog infrastructure can host, adapt, and accelerate GenAI for ambient intelligence, turning 6G networks into a distributed creative substrate rather than a passive transport layer.
\subsection{IoT Device Swarms for Distributed GenAI}
\label{subsec:iot-swarms}

IoT device swarms are large collections of heterogeneous, resource constrained devices that sense locally, communicate peer to peer or via an access point, and coordinate to accomplish learning and control tasks. In a 6G setting, such swarms complement edge and fog servers by pushing perception and inference closer to the physical world, thereby reducing backhaul load and latency while improving privacy by keeping raw data local. Recent advances in tiny machine learning demonstrate that meaningful inference can run directly on microcontrollers, enabling on site feature extraction and preliminary decision making within the swarm \cite{lin2020mcunet}. This local capability is important for ambient intelligence, because it allows devices to filter and compress observations into semantically rich signals before sharing them.

Training and adaptation across a swarm require communication efficient collaboration. Federated learning over wireless links is a natural fit, but the wireless channel creates bottlenecks and distortions that must be co designed with learning protocols. A foundational line of work analyzes federated optimization when devices share a fading multiple access channel, proposing digital and analog aggregation schemes, and proving convergence under realistic wireless impairments \cite{amiri2020flfading}. Building on the additive nature of the wireless medium, over the air aggregation can dramatically shorten model update time by summing local gradients in the air, which has been studied for broadband systems \cite{zhu2020broadband} and for extremely low bitrate regimes using one bit aggregation \cite{zhu2021onebit}. These results provide a path to scalable global model updates across dense swarms.

Swarms are often heterogeneous and intermittently connected, so orchestration is as important as the physical layer. Effective client selection and scheduling improve learning progress under limited energy and bandwidth budgets \cite{nishio2019fedcs}, and adaptive federated learning frameworks further align computation and communication to the time varying constraints of edge devices \cite{wang2019adaptivefl}. When memory limits prevent full models from fitting on devices, split learning and federated split variants partition models across devices and nearby servers, reducing local state while preserving privacy of raw data \cite{duan2022fedsplit}. Beyond star topologies, decentralized protocols over device to device graphs allow swarms to learn without a central server, which is attractive for non terrestrial or infrastructure poor scenarios \cite{xing2020decentralized}. Surveys that position federated learning as a key ingredient of 6G further emphasize these design points, connecting learning to integrated sensing, communication, and computation \cite{yang2022fl6g}.

Finally, swarm level coordination benefits from classic swarm intelligence mechanisms for robust, distributed optimization under strict energy and bandwidth budgets, which are common in IoT deployments \cite{abualigah2023swarmIoT}. Combined with physical layer aggregation techniques, including recent digital over the air designs, these algorithmic tools help swarms sustain low latency, privacy aware, and energy efficient distributed GenAI services at scale \cite{fan2021digitaloac}. As an optimization baseline for uplink access in dense swarms, exact solutions for power–subcarrier allocation with time-sharing in multicarrier NOMA provide a useful comparator to learning-based schedulers, clarifying the performance headroom under idealized assumptions \cite{10888922}

%============================================================
\subsection{Intelligent Reflecting Surfaces (IRS)}
\label{subsec:irs}

Intelligent reflecting surfaces (IRS) are engineered metasurfaces whose subwavelength elements impose programmable phase, amplitude, or polarization changes on incident waves, thereby shaping the wireless channel itself rather than only adapting transceivers \cite{wu2021tcom_tutorial,DiRenzo2020JSAC}. By turning walls, ceilings, and facades into controllable reflectors, IRS enables environment level control of propagation which has been shown to outperform the systems without IRS \cite{Umer2024CommLett}, which is valuable for ambient intelligence where perception and communication are tightly coupled. Tutorials and surveys have established IRS fundamentals, including electromagnetic and communication models, reflection control, hardware constraints, and deployment principles across single and multi surface topologies \cite{wu2021tcom_tutorial,DiRenzo2020JSAC,Pan2020RIS6G}. Empirical performance for STAR-RIS in CoMP-NOMA multi-cell settings further supports these gains \cite{Umer2023GCWkshps}. These works motivate IRS as a native component of 6G architecture, since it can improve coverage, spectral efficiency, energy efficiency, localization, and robustness in dynamic environments.

For distributed GenAI, IRS offers a physical layer lever to improve the data and compute pipeline. First, by increasing received signal strength and line of sight likelihood, IRS can lift the operating point of edge models, improving the accuracy and latency of semantic uplinks from sensors and wearables. Second, IRS can enhance the reliability of control channels that carry compressed intent or model updates, which reduces the need for retransmissions during federated or split generative learning. Third, IRS can aid joint communication and sensing, enabling richer environment perception that GenAI can fuse with historical context for ambient decision making. Recent overviews of RIS enabled integrated sensing and communication describe how surfaces can sculpt illumination and echo fields for more reliable sensing without dedicated radar hardware \cite{wang2023risenabledintegratedsensingcommunication}. For mobile deployments, deep reinforcement learning has been used to jointly optimize aerial-RIS trajectory and phase shifts in CoMP-NOMA networks, showing tangible gains in coverage shaping and link reliability under mobility \cite{10901709}.

Real world field trials have demonstrated the practicality of IRS at sub 6 GHz with large element counts, showing substantial power gains and sustained video throughput in non line of sight settings, while consuming only on the order of a watt at the surface controller \cite{Pei2021TCOM_field}. Such prototypes support the case for IRS as a deployable 6G enabler that can be co designed with edge computing and GenAI services in buildings and campuses.

A critical bottleneck for IRS assisted links is channel state information, particularly the cascaded base station, IRS, user channel. Conventional estimation requires long pilots that are mismatched to fast ambience changes. Machine learning methods, including model based and data driven optimization, have been surveyed for phase configuration and estimation under practical constraints \cite{Zhou2024COMST_Optim}. Growing evidence suggests that \emph{generative} models are especially well matched to this challenge. Conditional generative adversarial networks can map short pilot observations to high fidelity cascaded channel estimates, reducing overhead and improving robustness for sensing assisted links \cite{zhou2025cganris}. Diffusion models have been adapted to channel reconstruction by casting estimation as denoising with deterministic sampling, which improves stability and accuracy across signal to noise ratios and can be realized with lightweight U Net backbones suitable for the edge \cite{Wang2025RISDiffusion}. Complementing diffusion and GAN approaches, transformer-based distributed learning for downlink channel estimation in RIS-aided networks has also been demonstrated, enabling scalable PHY-layer inference under split-execution constraints \cite{11011078}.Beyond reconstruction, diffusion has also been paired with successive interference cancellation to jointly estimate channels and detect data in low-rank regimes, demonstrating reliable PHY-layer inference under practical constraints \cite{10888845}. Beyond estimation, generative channels for multi surface layouts open a path to scalable simulation and data augmentation for training ambient policies that must generalize across buildings, materials, and occupancy patterns \cite{DiRenzo2020JSAC}.

From a systems viewpoint, the synergy between IRS and distributed GenAI hinges on co design. Surfaces can expose semantic control interfaces to higher layers, for example, a small vocabulary of “illumination patterns” optimized offline by GenAI over digital twins, and called online as context evolves. Conversely, GenAI can provide surface aware scheduling, predicting mobility and blockage to pre configure phase maps that keep semantic links reliable. As IRS hardware matures toward hybrid or even partially active elements, and as learning centric optimization toolchains become standard \cite{Okogbaa2022Survey}, IRS will increasingly act as a \emph{programmable environment accelerator} for ambient intelligence, complementing edge and fog compute and the IoT substrate.
\subsection{Non terrestrial networks (NTNs)}
Non terrestrial networks extend ambient intelligent services beyond the footprint of terrestrial cells by adding connectivity from satellites, high altitude platform stations, and airborne relays. In practical deployments, NTNs must integrate with 3GPP New Radio so that user equipment, the core network, and radio resource control remain consistent across ground, air, and space segments. The 3GPP study and work items have specified key adaptations for waveforms, timing, and procedures in the presence of large propagation delays and high Doppler, thereby enabling direct access from user equipment to LEO, MEO, and GEO payloads as well as to HAPS nodes \cite{tr38821,lin2021space}. Recent surveys synthesize the architectural options and standardization trajectory from Release 17 toward 6G, highlighting service continuity, transparent and regenerative payloads, and orchestration across satellite constellations and terrestrial domains \cite{azari2022evolution,guidotti2020architectures}.

From the ambient intelligence perspective, NTNs matter for two reasons. First, they provide truly global reach for sensing, intent exchange, and actuation when the terrestrial fabric is unavailable or overloaded. Second, they unlock new placements for \emph{distributed GenAI}, including in orbit or stratospheric inference serving edge users over very wide areas. In NR NTN, Doppler, frequency offsets, and rapid line of sight dynamics require tailored physical layer and mobility designs. Measurements and analyses show that mobility and handover strategies must be rethought for fast moving LEO beams and for large differential delays across footprints \cite{juan2020leomobility,juan2022handover}. Link budget studies further clarify the tradeoffs among frequency band, antenna gain, and payload regeneration for both control and user plane continuity, which is essential when embedding predictive or generative models at the satellite edge \cite{guidotti2020linkbudget}. For integrated TN-NTN operation, hierarchical deep reinforcement learning has been proposed for intelligent spectrum sharing, coordinating cross-domain resources to sustain service continuity and spectral efficiency \cite{11016260}.

Massive IoT is a primary driver for NTN in 6G. Narrowband IoT over LEO exemplifies how scheduling and resource allocation must account for time varying visibility windows, differential Doppler limits, and signaling overheads that are far more stringent than in terrestrial cells. Uplink strategies that jointly consider coverage time, device demand, and Doppler constraints significantly improve delivery probability and energy efficiency, helping AmI workloads maintain telemetry and control under sparse opportunities \cite{kodheli2022nbiot}. Finally, NTNs can be paired with intelligent surfaces on aerial or spaceborne platforms to shape channels and alleviate blockage, creating energy aware, coverage adaptive links that are better suited to carry semantic messages and digital twin updates over large regions \cite{umer2025risntn}. At the interface of NTN and intelligent environments, deep reinforcement learning for active RIS-integrated TN-NTN has been shown to optimize multi-resource allocation policies, suggesting a path to coupling physical-layer reconfiguration with NTN scheduling \cite{10978440}.

In summary, NTNs provide the wide area backbone and placement diversity that distributed GenAI needs for ambient intelligence at planetary scale, while NR compliant procedures and mobility management keep the system interoperable with terrestrial 6G.
%============================================================
\section{Open Challenges and Research Directions}
Ambient intelligence at 6G scale requires generative models that are energy aware, privacy preserving, and interoperable with evolving standards. While Sections~3 and 4 demonstrated feasibility across models and enablers, production deployment reveals four cross cutting challenges: energy efficient on device training and adaptation, trustworthy synthetic data, federated generative learning under non identical distributions and weak links, and standardization tailored to AmI contexts. We discuss each in turn and outline concrete research directions.

\subsection{Energy efficient on device training and adaptation}
Edge and device scale learning will be central to AmI, yet sustained training and continual fine tuning can exhaust energy budgets. Experience from deep learning shows that compute and energy scale sharply with model size and search, motivating algorithmic and system co design \cite{strubell2019energy}. Promising directions include \emph{once for all} supernet training with hardware aware specialization for local constraints \cite{cai2020onceforall}, efficient low rank adaptation to avoid full model updates, and quantization aware fine tuning such as QLoRA for resource constrained personalization \cite{hu2021lora,dettmers2023qlora}. On the systems side, dynamic partitioning and offload policies should consider \emph{training} phases, not only inference, coordinating gradient steps across device, edge, and cloud with energy as a first class objective. Benchmarks that report wall clock, energy, and accuracy for common AmI tasks are needed to make progress comparable across platforms.

\subsection{Trustworthy synthetic data}
Generative models can amplify scarce measurements, but they also introduce privacy, memorization, and bias risks. Membership inference and extraction attacks have demonstrated leakage from generative models, including diffusion models \cite{hayes2018loganmembershipinferenceattacks,carlini2023extractingtrainingdatadiffusion}. Differentially private learning provides a principled defense, but introduces accuracy and utility tradeoffs that must be characterized for AmI sensing and control loops \cite{abadi2016deepdp}. Open problems include formal privacy accounting for \emph{conditional} generators used with side information, distribution shift detection when synthetic data drives downstream decisions, and standardized utility–privacy evaluation protocols that reflect AmI workloads rather than generic image benchmarks.

\subsection{Federated generative learning under wireless constraints}
Ambient deployments are naturally federated, with non-identical users, intermittent links, and strict communication budgets. Classical results on communication efficient federated optimization \cite{mcmahan2017communication} and variance controlled updates \cite{karimireddy2020scaffold} inform the design of \emph{generative} FL, but open issues remain: mode coverage across clients with disjoint supports, stragglers and drift, and robustness to partial participation. Stabilizing diffusion and adversarial generators under federated heterogeneity, combining split learning with federated aggregation to reduce device memory, and leveraging over the air aggregation where appropriate are fertile directions. Practical recipes that align local steps, compression, and personalization with radio budgets are needed to move beyond small scale demos \cite{li2020federatedoptimizationheterogeneousnetworks}.

\subsection{AmI specific standardization and governance}
Ambient intelligence spans devices, radio, compute, and data governance. Interoperability requires interfaces for model lifecycle, telemetry, and control that align with emerging standards. ITU T’s architectural framework for machine learning in future networks defines generic lifecycle and management hooks \cite{itutY3172}, and the O RAN Alliance specifies AI and ML workflows in the RAN for data collection, model hosting, and policy control \cite{oran2022aiml}. For safety, the NIST AI Risk Management Framework provides developer and operator guidance on mapping, measuring, and managing model risks across the lifecycle \cite{nistrmf2023}. An AmI profile of these efforts is still missing. Concretely, research should define semantic communication interfaces for intent and context, auditing hooks for synthetic data pipelines, and conformance tests that couple wireless performance with privacy, robustness, and energy criteria.
%============================================================
\section{Conclusion}
Ambient intelligence envisions environments that can perceive context, infer intent, and act autonomously in service of people. This chapter argued that realizing such environments at global scale requires sixth generation networks that are not only faster and more reliable, but are also natively perceptive, predictive, and controllable. Within this vision, generative artificial intelligence emerges as the creative core that enables sensing, reasoning, and actuation to operate coherently across the device–edge–cloud continuum.

We surveyed the algorithmic foundations of modern generative modeling, including adversarial learning, variational inference, diffusion, and transformer based generators, and we highlighted why each family is complementary for ambient intelligence. Adversarial learning offers high fidelity synthesis and scenario stress testing, variational methods provide structured latent spaces and principled uncertainty handling for compression and estimation, diffusion models deliver stable and diverse generation with precise conditional control, and generative transformers supply general purpose priors for planning, tool use, and multimodal reasoning. Together, these models allow ambient systems to fill data gaps, simulate rare events, compress semantics into compact messages, and forecast network and user states for proactive control.

We then connected these capabilities to 6G enablers that can host or accelerate distributed generative intelligence. Edge and fog computing place learning close to data to reduce delay and protect privacy, IoT device swarms widen the sensing and actuation surface while supporting collaborative learning, intelligent reflecting surfaces shape channels to raise reliability and sensing quality, and non terrestrial networks extend coverage and placement options beyond terrestrial cells. Co designed with generative models, these enablers transform the network from a passive conduit into an adaptive substrate that sustains ultra reliable, low latency, and context aware services.

Despite this promise, we identified cross cutting challenges that must be addressed before ambient intelligence becomes pervasive. Energy efficient training and adaptation at devices and edges are necessary to respect tight power budgets. Trustworthy synthetic data pipelines must guard against memorization, leakage, and bias while preserving utility for downstream decisions. Federated generative learning has to operate under non identical data, intermittent links, and limited communication, calling for algorithms that align computation, compression, and radio constraints. Finally, ambient intelligence specific standardization and governance are needed to define interoperable interfaces for model lifecycle, telemetry, safety, privacy, and conformance, so that deployments remain auditable and trustworthy.

A practical roadmap follows from these observations. First, define semantic communication interfaces so that intent, context, and uncertainty can be exchanged efficiently across devices, reflectors, edges, and satellites. Second, co design surfaces, radios, and learning policies within digital twins that incorporate generative channel and workload models, enabling safe what if exploration and data efficient optimization. Third, adopt parameter efficient adaptation and fast generators to meet device and edge limits, while instrumenting systems to report accuracy, energy, latency, and privacy metrics in a comparable way. Fourth, align emerging standards with ambient intelligence profiles, ensuring that safety and governance requirements travel with models and data across administrative and geographic boundaries.

In closing, generative artificial intelligence is not a peripheral addition to sixth generation networks. It is the mechanism that lets networks perceive, predict, and act in context, turning connectivity into a distributed intelligence fabric. By integrating generative models with edge and fog computing, IoT swarms, intelligent reflecting surfaces, and non terrestrial networks, 6G systems can transition from reactive data pipes to proactive ambient companions that improve safety, efficiency, and human experience across everyday spaces.
%============================================================

%============================================================
\bibliographystyle{elsarticle-num}
% if main.bib lives in  /references/main.bib
\bibliography{references/main}

%============================================================
\end{document}